\documentclass[]{pasj01}
\draft
\usepackage{multirow}

\begin{document}
\Received{}
\Accepted{}

\title{AKARI far-infrared maps of the zodiacal dust bands}


 \author{%
   Takafumi \textsc{Ootsubo}\altaffilmark{1},
   Yasuo \textsc{Doi}\altaffilmark{1},
   Satoshi \textsc{Takita}\altaffilmark{2},
   Takao \textsc{Nakagawa}\altaffilmark{2},
   Mitsunobu \textsc{Kawada}\altaffilmark{2},
   Yoshimi \textsc{Kitamura}\altaffilmark{2},
   Shuji \textsc{Matsuura}\altaffilmark{3},
   Fumihiko \textsc{Usui}\altaffilmark{4}, and
   Ko \textsc{Arimatsu}\altaffilmark{5}
        }
 \altaffiltext{1}{Department of Earth Science and Astronomy, Graduate School of Arts and Sciences, 
 The University of Tokyo, Komaba 3-8-1, Meguro, Tokyo 153-8902, Japan}
 \email{ootsubo@ea.c.u-tokyo.ac.jp}
 \altaffiltext{2}{Institute of Space and Astronautical Science, Japan Aerospace Exploration Agency, 
 3-1-1 Yoshinodai, Chuo, Sagamihara, Kanagawa 252-5210, Japan}
 \altaffiltext{3}{Department of Physics, Graduate School of Science and Technology, Kwansei Gakuin University,
 2-1 Gakuen, Sanda, 669-1337, Japan}
 \altaffiltext{4}{Department of Astronomy, Graduate School of Science, The University of Tokyo, 
 7-3-1 Hongo, Bunkyo, Tokyo 113-0033, Japan}
 \altaffiltext{5}{National Astronomical Observatory of Japan, 
 2-21-1 Osawa, Mitaka, Tokyo 181-8588, Japan}


\maketitle

\begin{abstract}
Zodiacal emission is thermal emission from interplanetary dust. Its contribution to the sky brightness is non-negligible in the region near the ecliptic plane, even in the far-infrared (far-IR) wavelength regime.
We analyse zodiacal emission observed by the AKARI far-IR all-sky survey, which covers 97\% of the entire sky at arcminute-scale resolution in four photometric bands, with central wavelengths of 65, 90, 140, and 160~$\mu$m.
AKARI detected small-scale structures in the zodiacal dust cloud, including the asteroidal dust bands and the circumsolar ring, at far-IR wavelengths.
Although the smooth component of the zodiacal emission structure in the far-IR sky can be reproduced well by models based on existing far-IR observations, previous zodiacal emission models have discrepancies in the small-scale structures compared with observations. 
We investigate the geometry of the small-scale dust-band structures in the AKARI far-IR all-sky maps and construct template maps of the asteroidal dust bands and the circumsolar ring components based on the AKARI far-IR maps.
In the maps, $\pm \timeform{1.4D}$, $\pm \timeform{2.1D}$ and $\pm \timeform{10D}$ asteroidal dust-band structures are detected in the 65~$\mu$m and 90~$\mu$m bands. 
A possible $\pm \timeform{17D}$ band may also have been detected. 
No evident dust-band structures are identified in either the 140~$\mu$m or the 160~$\mu$m bands.
By subtracting the dust-band templates constructed in this paper, we can achieve a similar level of flux calibration of the AKARI far-IR all-sky maps in the $|\beta| < 40\degree$ region to that in the region for $|\beta| > 40\degree$.
\end{abstract}

\section{Introduction}
%
%
Zodiacal emission (ZE) is thermal emission from interplanetary dust (IPD). It dominates the diffuse radiation in the mid- to far-infrared (IR) wavelength regime. 
Since the dust located at around 1 astronomical unit (a.u.) from the Sun has a thermal equilibrium temperature of $\sim 280$ K, the ZE exhibits a spectral peak around 10--20~\micron, which dominates the mid-IR brightness of the diffuse emission in the sky.
Even in the far-IR regime, the ZE contribution is non-negligible at high Galactic latitudes, where Galactic radiation is relatively weak. 
IPD originates mainly from comets and asteroids, and it is distributed across the solar system. 
The zodiacal dust cloud is characterized by a relatively smooth distribution, resulting in the dominant diffuse emission. 
However, observations with the Infrared Astronomical Satellite (IRAS) revealed that small-scale structures (on scales of a few to a few tens of degrees) are present in the ZE distribution, such as those tracing the asteroidal dust bands and a circumsolar resonance ring \citep{Low84, Dermott84}.

%
%
The zodiacal dust bands were first discovered with IRAS in its 12~$\mu$m to 100~$\mu$m data \citep{Low84, Neugebauer84}.
They have subsequently been confirmed by the Diffuse Infrared Background Experiment (DIRBE) onboard the Cosmic Background Explorer (COBE), as well as by ground-based observations \citep{Spiesman95, Reach97, Ishiguro99a, Ishiguro99b}. 
They provide a link between the asteroidal source of the IPD and the present distribution of the zodiacal cloud. 
There are three major band pairs 
among the dust bands that form small-scale latitudinal features in the ZE.
They are thought to be produced by thermal emission from three pairs of circumsolar dust rings near inclinations
with respect to the invariant plane of the planets, $i_{\rm p} = \pm \timeform{1.4D}$, $\pm \timeform{2.1D}$, and $\pm \timeform{9.3D}$ \citep{Sykes88, Sykes90, Grogan01}. 
The two central bands (at $\pm \timeform{1.4D}$ and $\pm \timeform{2.1D}$)
are blended in low-resolution data (see e.g., \cite{Reach97}), and high-resolution maps are required to identify and distinguish the band pairs.
It has been proposed that recent disruption events among multi-kilometer bodies in the main asteroid belt, which would have occurred within the last several million years, may be major supply sources of dust particles, and they would produce edge-brightened toroidal dust distributions (referred to as the ``catastrophic'' or ``non-equilibrium'' model; \cite{Sykes86, Sykes88}).
Recent detailed studies \citep{Nesvorny02, Nesvorny03, Nesvorny06a, Nesvorny08} have provided evidence supporting the disruption model as the possible origin of the asteroidal dust bands.  
\citet{Nesvorny03} proposed that the dust band characterized by an inclination of $\pm \timeform{9.3D}$ may have originated from the Veritas asteroid family, at a distance of 3.17 a.u. from the Sun, which formed through collisional disruption $8.3 \pm 0.5$ Myr ago, while the $\pm \timeform{2.1D}$ band may have originated from the Karin family, which formed $5.75 \pm 0.05$ Myr ago and is located inside the Koronis asteroid family at a solar distance of 2.87 a.u. \citep{Nesvorny02, Nesvorny03, Nesvorny06b}. 
\citet{Nesvorny08} also proposed a recent breakup of the new candidate Beagle asteroid family as the source of the $\pm \timeform{1.4D}$ band.

%
%
Dust from the asteroids and comets, which continues to spiral in towards the inner Solar System, is subject to gravitational perturbations by the planets. The presence of a circumsolar ring (i.e., a ring structure of dust originating from the outer Solar System that is resonantly trapped by the Earth in orbits near 1 a.u. from the Sun) has been proposed theoretically (e.g., \cite{Jackson89}). 
The sky brightness near the ecliptic plane, as observed with IRAS, is brighter in the Earth-trailing than in the leading direction \citep{Dermott94}.
\citet{Dermott94} showed, based on numerical simulations, that this brightness asymmetry is caused by a trailing blob of dust particles trapped in mean-motion resonances (MMRs) with the Earth. 
This Earth's resonance structure has been confirmed using COBE/DIRBE observations \citep{Reach95}. 
The existence of a similar circumsolar ring at the orbit of Venus has also been confirmed based on the imaging data of the STEREO mission \citep{Jones13}.
%
%
In addition to these small-scale structures on scales of a few to a few tens of degrees, finer-scale features associated with the orbits of periodic comets, the so-called cometary dust trails, have also been observed in the IRAS survey data \citep{Sykes86s, Sykes92, Reach00}.

%
%
To determine the brightness distribution of the ZE and the spatial structure of the IPD, full-sky surveys obtained with IR satellites are particularly powerful (e.g., \cite{Hauser84, Reach92, Reach97, Fixen02, Pyo10, Planck14-14}). 
Subsequent to IRAS, COBE/DIRBE conducted an all-sky survey with a wider wavelength coverage than IRAS \citep{Hauser98}.
Many efforts have been devoted to describing the structure of the zodiacal dust cloud, and a number of models have been developed using IR satellite data, including the IRAS and COBE/DIRBE surveys
\citep{Good86, RR90, RR91, Jones93, Vrtilek95, Kelsall98, Wright98, RR13}.
The most commonly used ZE models are based on the DIRBE data (e.g., \cite{Kelsall98}: hereafter the Kelsall model; \cite{Wright98}: hereafter the Wright model).
\citet{Gorjian00} investigated the DIRBE ZE model in depth and modified the parameters of the Wright model (\cite{Wright98}).
The Kelsall and the Wright/Gorjian models explain the IPD cloud complex with the following three components:
a smooth cloud, asteroidal dust bands and an MMR component.
In the Kelsall model, the MMR component is further divided into a circumsolar ring and a blob trailing the Earth \citep{Kelsall98}.
For each of the three components, parameterised functions have been introduced to describe the spatial distribution of the dust number density in heliocentric coordinates.

%
%
These models, however, have limitations in their ability to reproduce small-scale structures.
The COBE/DIRBE team provided Zodi-Subtracted Mission Average (ZSMA) map data, which were created by subtracting the ZE using the Kelsall model week-by-week and averaging the residual intensity values. Small-scale residual ZE structures near the ecliptic plane can be seen clearly in the ZSMA maps, in particular at mid-IR wavelengths (e.g., \cite{Kelsall98}).
It is therefore necessary to construct a ZE model based on IR-satellite data with higher spatial resolution.

%
%
Following the completion of the IRAS and COBE/DIRBE missions, a few satellites have since conducted full-sky surveys in the IR regimes, including AKARI and the Wide-field Infrared Survey Explorer (WISE; \cite{Wright10}). 
Table \ref{tab:irsurveys} offers a comparison of major full-sky mid- and far-IR satellite surveys.
The Japanese satellite AKARI, a satellite dedicated to IR astronomical observations, was the third mission to survey the whole sky at mid- and far-IR wavelengths \citep{Murakami07}.
Although most of the smooth ZE structure seen in the AKARI mid- and far-IR images can be reproduced well with the Kelsall or the Wright/Gorjian models, there are discrepancies on small scales \citep{Pyo10, Kondo16}.
In particular, the intensities and the ecliptic latitudes of the peak positions of the asteroidal dust bands and the MMRs cannot be precisely reproduced with these models \citep{Pyo10, Doi15, Kondo16}.
The DIRBE model cannot reproduce the real fine structure of the asteroidal dust-band components, because DIRBE operated with a large $42\arcmin \times 42\arcmin$ beam, and only a small fraction of the DIRBE data were used to construct the DIRBE ZE model \citep{Kelsall98, Wright98}.
\citet{Pyo10} also reported an inconsistency of 20\% in the intensity of the ring component between the AKARI observations and the Kelsall model prediction in the mid-IR regime.

Since the ZE is the nearest source of diffuse emission to the Earth and an important foreground component for studies of Galactic and extragalactic emission at IR wavelengths, it is extremely important to understand the nature of the zodiacal dust cloud, not only for Solar System science.
It is crucial to precisely evaluate the contribution of the ZE to the IR sky brightness.
The diffuse emission in the far-IR regime is, for instance, a direct measure of the column density of the interstellar dust,
and a full-sky far-IR map can also be used as a measure of the extinction for the emission from extragalactic objects. 
A careful treatment of the ZE is also required for the construction of extinction maps \citep{SFD98, FDS99, Meisner14, Planck14-11}.

We performed an all-sky survey with the AKARI satellite \citep{Murakami07}. The all-sky far-IR map data have been publicly released \citep{Doi15}. 
AKARI covers wavelengths longer than 100~$\micron$ and has a higher spatial resolution (1$\arcmin$--1.5$\arcmin$) than IRAS.
It is important to obtain a precise estimate of the ZE contribution to the far-IR sky and provide a ZE template and ZE-subtracted far-IR maps based on the AKARI data.  
This paper is part of a series in which we present AKARI far-IR diffuse emission map data. 
\citet{Doi15} provide detailed descriptions of both the AKARI far-IR all-sky survey and the data-processing scheme applied to the all-sky maps.
\citet{Takita15} report on the calibration process and quality of the AKARI far-IR all-sky maps.
In this paper, we evaluate the residual ZE contribution, in particular of the zodiacal dust bands, in the publicly released AKARI far-IR maps and discuss the AKARI ZE-distribution template maps. 
We extract the structure of the zodiacal dust bands from the AKARI far-IR maps and briefly discuss their observed properties.
The main goal of this study is to provide ZE template maps and ZE-subtracted far-IR maps based on the AKARI far-IR all-sky survey data\footnote{ZE template maps will be provided at http://www.ir.isas.jaxa.jp/AKARI/Archive/}. 
Detailed analysis of the dynamics of the dust bands, much smaller-scale structures (such as cometary dust trails) and the construction of a three-dimensional ZE model (such as the Kelsall model) are beyond the scope of this study but will be explored in forthcoming papers.


\section{Observations and Data Sets}

\subsection{AKARI observations and far-IR all-sky maps}
%
%
AKARI was designed as an all-sky survey mission in the mid- and far-IR wavelength regime \citep{Murakami07}.
The satellite was equipped with a cryogenically cooled telescope of 68.5~cm aperture diameter and two scientific instruments,
the Far-Infrared Surveyor (FIS; \cite{Kawada07}) and
the Infrared Camera (IRC; \cite{Onaka07}).
AKARI was launched in 2006 February and the all-sky survey was performed during the period 2006 April to 2007 August, which was the cold operation phase of the satellite, when it was cooled with liquid helium cryogen.
Its 16-month cryogenic mission lifetime is thus by far the longest among the far-IR survey missions.
The all-sky survey observations at far-IR wavelengths were performed by employing the FIS instrument \citep{Kawada07}. Four photometric bands were used: the
N60 band (50--80~\micron, centred at 65~\micron), the
WIDE-S band (50--110~\micron, centred at 90~\micron), the
WIDE-L band (110--180~\micron, centred at 140~\micron) and the
N160 band (140--180~\micron, centred at 160~\micron).
The AKARI far-IR all-sky survey maps\footnote{The data can be retrieved from http://www.ir.isas.jaxa.jp/AKARI/Archive/}, which have a better spatial resolution than the IRAS maps, were recently released publicly \citep{Doi15}.

\begin{table}
  \caption{Comparison of IR satellite survey observations (sky coverage $>90$\%) at mid- and far-IR wavelengths.}\label{tab:irsurveys}
    \begin{tabular}{lcccc}
      \hline
      Satellite/Instrument & Observation dates & Wavelength ($\micron$) & Spatial resolution  \\
     \hline
      AKARI/FIS$^{a}$  & 2006 May--2007 Aug & 65, 90, 140, 160         & $\sim 60\arcsec$--90$\arcsec$  \\
      AKARI/IRC$^{b}$  & 2006 May--2007 Aug & 9, 18                    & $\sim 6\arcsec$  \\
      IRAS$^{c}$       & 1983 Feb--1983 Nov & 12, 25, 60, 100          & $\sim 5\arcmin$  \\
      COBE/DIRBE$^{d}$ & 1989 Dec--1990 Sep & 1.25, 2.2, 3.5, 4.9, 12, & $\sim \timeform{0.7D}$  \\
                       &                    & 25, 60, 100, 140, 240    & & \\             
      WISE$^{e}$       & 2010 Jan--2010 Aug & 3.4, 4.6, 12, 22         & $\sim 6$--12$\arcsec$  \\
	\hline
    \end{tabular}
	\begin{tabnote}
	$^{a}$ \citet{Doi15}. \\ 
    $^{b}$ \citet{Ishihara10}. \\
    $^{c}$ Explanatory Supplement to the IRAS Sky Survey Atlas (http://irsa.ipac.caltech.edu/IRASdocs/issa.exp.sup/). \\
    $^{d}$ DIRBE Explanatory Supplement (http://lambda.gsfc.nasa.gov/product/cobe/dirbe\_exsup.cfm). \\
    $^{e}$ The fully cryogenic survey phase. \citet{WISE12}. 
    \end{tabnote}
\end{table}

%
%
The observed far-IR sky brightness consists of the ZE, the integrated light from the Galactic components (stars and emission from the interstellar medium) and extragalactic emission.
Since AKARI far-IR all-sky map data are processed mainly to study the Galactic and extragalactic objects, the ZE components in the released maps have been subtracted during data processing.
As mentioned previously, although we tried to subtract the ZE from the AKARI far-IR data by applying the DIRBE ZE models \citep{Kelsall98, Wright98, Gorjian00}, we found that there are discrepancies on small scales between the model estimates and the observed data, especially in relation to the asteroidal dust bands and the MMRs.
Under these circumstances, we only subtracted the smooth component of the ZE, based on the Gorjian model \citep{Gorjian00}, from the publicly released AKARI far-IR maps (for more details, see \cite{Doi15}).
We have not yet subtracted the asteroidal dust-band and the MMR components from the data. Consequently, contributions from these components remain present in the images, and this is recognisable in the shorter-wavelength bands (N60 and WIDE-S).
Next, we investigate the residual ZE contribution to the AKARI far-IR all-sky maps and construct a ZE template map, including the asteroidal dust bands and the MMRs.

\subsection{Leading/trailing maps and data reduction}
%
%
The AKARI orbit and observation strategy are described in detail by \citet{Murakami07} and \citet{Doi15}.
AKARI orbited the Earth in a Sun-synchronous polar orbit with a period of about 100 min.
During each orbit, the satellite scanned the sky in the Earth-leading and trailing directions. 
The AKARI orbit processed $\sim 1\degree$ per day, and the AKARI scans were obtained at an approximately constant solar elongation of $90 \pm 1\degree$.
A given longitude, observed in the Earth-leading direction, was re-observed in the Earth-trailing direction from a different position 6 months later.
Using both the leading and trailing scans, AKARI achieved almost complete sky coverage in 6 months, although some regions are affected by the South Atlantic Anomaly (an anomalously higher-density region of solar protons) and the Moon \citep{Doi15}.
AKARI observed $97\%$ of the sky twice or more during the 16 months of its cryogenic mission lifetime, and the full data set from all 16 months was used for the construction of the publicly released AKARI maps \citep{Doi15}.  
Therefore, most sky regions in the AKARI image were processed with data from different satellite (Earth) positions and lines of sight (both in the leading and trailing directions). 

\begin{figure*}  
   \begin{center}
    \includegraphics[width=17cm]{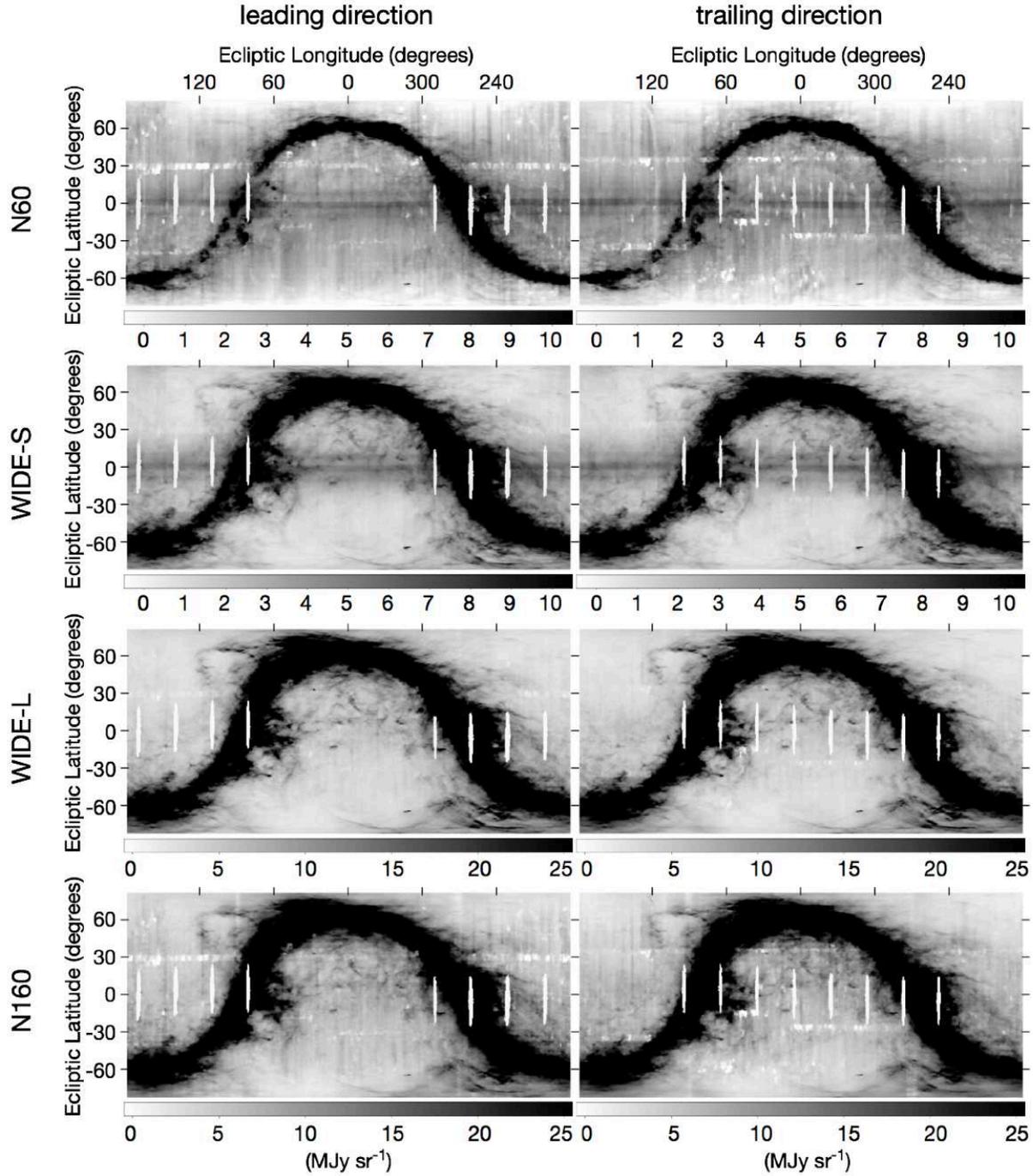} %
   \end{center}
\caption{AKARI far-IR all-sky surface brightness maps in four bands for the Earth-leading (left-hand panels) and trailing (right-hand panels) directions; from top to bottom: N60, WIDE-S, WIDE-L and N160.
The maps are rendered in Cartesian projection in the ecliptic coordinate system. The grey scale represents the surface brightness in units of MJy~sr$^{-1}$, as shown along the bottom bar. The map centres correspond to the sky position (0$\degree$, 0$\degree$) in ecliptic coordinates. 
The grid ticks are shown at longitudes 120$\degree$, 60$\degree$, 0$\degree$, 300$\degree$, and 240$\degree$ from left to right, and at latitudes -60$\degree$, -30$\degree$, 0$\degree$, 30$\degree$, and 60$\degree$ from bottom to top for each panel. Eight vertically long regions near the ecliptic plane contaminated by Moon light are left blank in each map. Horizontal whitish pattern around $\beta \sim \pm 30\degree$ are regions affected by the satellite maneuver for pointed observations.}
\label{fig:ltmaps}
\end{figure*}

The IPD cloud is asymmetric with respect to the Earth, and the observed ZE brightness changes with satellite position.
AKARI observed the Galactic and extragalactic components along almost the same lines of sight, whereas dust in the Solar System was observed along completely different lines of sight from opposite positions, 6 months apart along the Earth's orbit.
To investigate the ZE component, we therefore divided and reprocessed each of the released AKARI far-IR maps for all four bands into two separate maps, according to whether they were observed in the leading or trailing directions.
The original AKARI maps have a pixel scale of 15$\arcsec$ and a spatial resolution (i.e., the full width at half maximum of the point-spread function) of 1--1.5$\arcmin$ \citep{Takita15}.
The asteroidal dust bands and the MMR components exhibit much more extensive spatial structures, covering a few and a few tens of degrees along the latitudinal direction, respectively.
To reduce the noise and contributions from both much smaller-scale structures and point sources, we reconstructed the maps with a resolution of \timeform{0.2D}. 
We use these maps, in the Earth-leading and trailing directions, to extract the structure of the ZE distribution in the following analysis.
The processed sky-brightness maps in the four FIS bands 
are shown in figure \ref{fig:ltmaps} for both the leading and trailing directions.
Other IR-satellite survey missions (e.g., IRAS, COBE/DIRBE or WISE) lack certain regions in longitude ($\sim 15$\%) in their leading/trailing maps, because their cryogenic mission lifetimes were shorter than 1 year. 
The AKARI survey data cover all ecliptic longitudes ($\lambda = 0\degree$--360$\degree$), except for the small regions affected by the Moon, both in the leading and trailing maps within the satellite's 16-month cryogenic mission period.
By using the AKARI leading/trailing maps, we can investigate the continuous dust-band structures along all ecliptic longitudes.
%
%

Asteroidal dust bands appear as pairs of equally spaced parallel bands that are oriented along the ecliptic longitude direction in a sinusoidal fashion above and below the ecliptic plane. 
The continuous, smooth distribution of the dust bands and the MMR component along the ecliptic plane 
can be seen in the N60 and WIDE-S maps (see figure \ref{fig:ltmaps}).
In the ``snap-shot image'' of the DIRBE weekly sky map covering the wide solar elongation angle of 60$\degree$--120$\degree$, the MMR component in the Earth's orbit shows an elliptical structure. The full-width at half-maximum brightness of the trailing blob is 30$\degree$ in latitude and 15$\degree$ in longitude \citep{Reach95}. AKARI always scans the central part of the MMR component at the solar elongation of $\sim 90\degree$. As a consequence, the superposition of the MMR component in the AKARI all-sky map appear as a horizontal faint band structure of $\sim 30\degree$ width in latitude oriented along the ecliptic longitude direction.
Note that the ZE smooth cloud component has been subtracted from the AKARI images.
To reduce the effects of non-Solar System contributions to the sky brightness and to extract the dust-band distribution, we constructed difference maps using the WIDE-S and WIDE-L bands.
The two wide bands (WIDE-S and WIDE-L) yield higher-quality images because they had a broader wavelength coverage than the two narrow bands (N60 and N160).

\subsection{Difference maps based on the WIDE-S and WIDE-L data}
%
%
Since Galactic emission is strong in the far-IR regime, extraction of the faint dust bands from the far-IR maps is not a simple task. 
To measure the dust-band structure precisely, image-enhancing techniques were used.
Fourier filtering is a commonly used technique to extract structures from an image (e.g., \cite{Sykes88, Reach97}).
However, it is likely that smooth and broad components of the dust-band structures are removed by Fourier filtering.
Since the main aim of this paper is to construct ZE templates that can be subtracted from the AKARI far-IR images, we adopted the following approach to deduce the intensities and positions of the small-scale ZE structures. 

\begin{figure}
 \begin{center}
   \includegraphics[width=10cm]{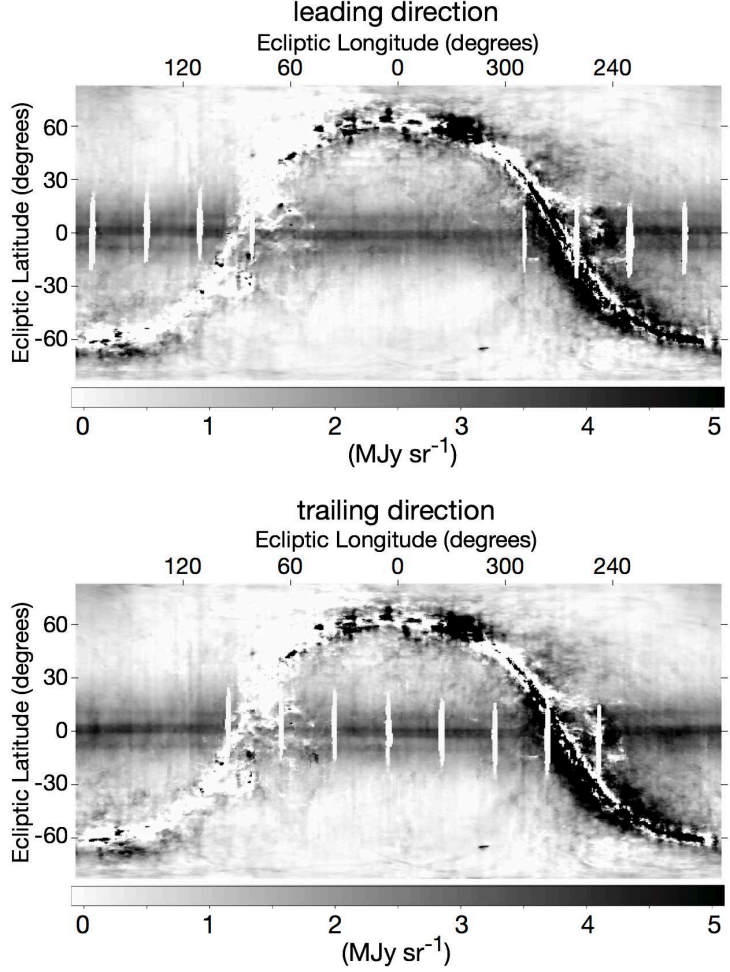} %
 \end{center}
\caption{Difference maps based on comparison of the WIDE-S and WIDE-L bands, $I_\mathrm{WIDE-S}-0.30 \times I_\mathrm{WIDE-L}$, in the leading (top) and trailing (bottom) directions. The maps are rendered in Cartesian projection in the ecliptic coordinate system, as in figure \ref{fig:ltmaps}. There can be seen residual Galactic emission near the Galactic plane (black region), which is attributed to the emission from the Galactic dust that has the temperature different from $T = 18.5$~K.}
\label{fig:diffmap}
\end{figure}

The WIDE-S (90~$\micron$) maps include both IPD and interstellar dust emission, while the WIDE-L (140~$\micron$) maps, which cover longer wavelengths than their WIDE-S counterparts, mainly detect interstellar dust emission (IR Galactic cirrus). 
The dust grains in the asteroidal dust bands spiral inward from the main asteroid belt (located at a distance from the Sun of $\sim 2$--3~a.u.) to near the Earth's orbit and have a temperature of around 150--300~K.
MMR dust, at $\sim 1$~au, has a temperature of around 280~K.
Such dust exhibits a spectral peak at $\lambda < 50~\micron$ \citep{Spiesman95, Reach97}.  
The ZE contribution to the WIDE-L (140~$\micron$) map is much smaller than that to the WIDE-S (90~$\micron$) map.
To reduce the contribution from Galactic dust and extract the small-scale ZE structures,
we subtracted 0.30 times the WIDE-L intensity ($0.30 \times I_\mathrm{WIDE-L}$) from the WIDE-S intensity ($I_\mathrm{WIDE-S}$) for both the leading and trailing maps. 
We derived the factor of 0.30 as follows:
\begin{equation}
\frac{\int_\mathrm{WIDE-S} B_\nu(18.5~\mathrm{K})~(\nu/\nu_{0})^{2}~R_\mathrm{WIDE-S}(\nu)~ \mathrm{d}\nu
/ \int_\mathrm{WIDE-S} R_\mathrm{WIDE-S}(\nu)~ \mathrm{d}\nu}
{\int_\mathrm{WIDE-L} B_\nu(18.5~\mathrm{K})~(\nu/\nu_{0})^{2}~R_\mathrm{WIDE-L}(\nu)~ \mathrm{d}\nu
/ \int_\mathrm{WIDE-L} R_\mathrm{WIDE-L}(\nu)~ \mathrm{d}\nu} \sim 0.30,
\end{equation}
where $B_\nu(18.5~\mathrm{K})$ is the Planck function at a temperature $T = 18.5$~K, $R_\mathrm{WIDE-S}(\nu)$ and $R_\mathrm{WIDE-L}(\nu)$ are the relative spectral responses of the WIDE-S and WIDE-L bands, respectively \citep{Doi15}, and $\nu_{0}=3000$~GHz is the reference frequency corresponding to $\lambda \simeq 100~\micron$.
Based on previous studies (e.g., \cite{SFD98, Planck14-11}), we assumed that the temperature of the interstellar dust is $T \sim 18.5$~K and the emissivity of the interstellar dust has a frequency (wavelength) dependence $(\nu/\nu_{0})^{2}$ at wavelengths longwards of 100~$\micron$.
In these ($I_\mathrm{WIDE-S}-0.30 \times I_\mathrm{WIDE-L}$) maps (see figure \ref{fig:diffmap}; hereafter the ``difference maps''),
we can see prominent sinusoidal dust bands near the ecliptic plane and at $\beta \sim \pm 10\degree$.
These images are used to determine the locations and brightnesses of the dust bands.

\section{Results: Zodiacal dust-band structure in the AKARI far-IR maps}

\subsection{Geometry of asteroidal dust bands and MMRs}

\begin{figure}
 \begin{center}
  \includegraphics[width=17cm]{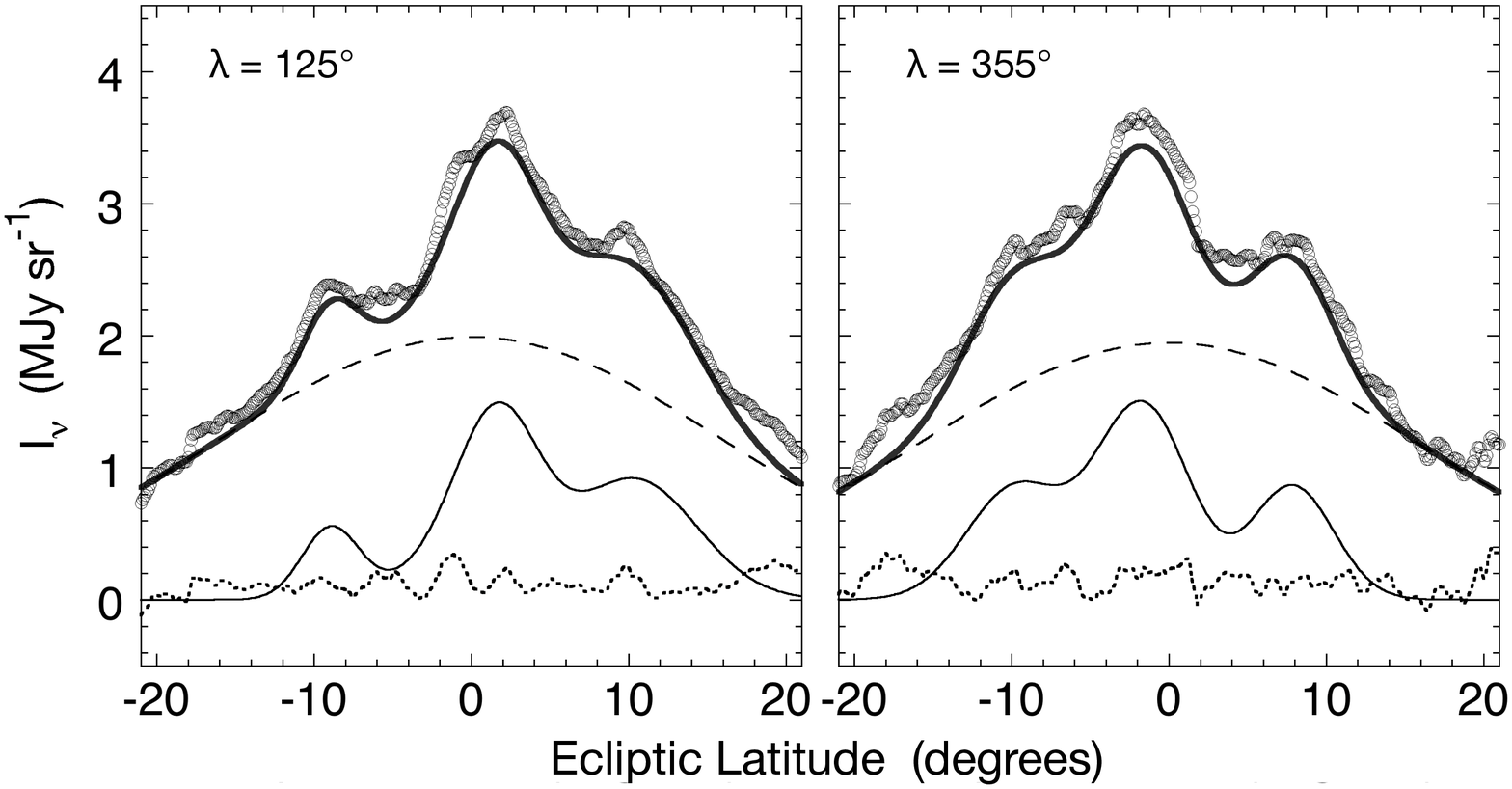} %
 \end{center}
\caption{Latitudinal profiles taken from the difference map at $\lambda = 125\degree$ and $355\degree$ in the Earth-leading direction (open circles) and a five-Gaussian fit (thick solid line), together with the circumsolar-ring (dashed line) and dust-band components (thin solid line). Dotted line: residual emission.} 
\label{fig:latprofile}
\end{figure}

To measure the latitudes of the dust bands and the structure of the circumsolar ring, we fitted the latitudinal profiles at $|\beta| < 25\degree$ with Gaussian functions. 
Bands at $\pm \timeform{1.4D}$ and $\pm \timeform{2.1D}$ were identified separately in the finer-scale analysis of the $\timeform{0.05D}$ resolution map using the 25~$\micron$ IRAS Infrared Sky Survey Atlas data, which were observed at a wavelength near the peak wavelength of the ZE spectrum \citep{Reach97}.
However, in the $\timeform{0.2D}$ resolution AKARI maps at much longer wavelengths, these two bands are apparently almost blended and look to be a single band. Hereafter, we refer to this pair as the $\pm \timeform{1.4D}/\timeform{2.1D}$ band.
Five-Gaussian fits (two Gaussians each for the $\pm\timeform{1.4D}/\timeform{2.1D}$ and $\pm \timeform{10D}$ dust-band pairs and one for the circumsolar ring) were performed to latitudinal profiles every 1$\degree$ in longitude, for the leading and trailing maps in figure \ref{fig:diffmap}. 
Although we reduced the Galactic cirrus contribution using the difference maps, a Galactic-emission component remains present in the maps, as seen in figure \ref{fig:diffmap}. 
No $|\beta| < 25\degree$ scans passing within 15$\degree$ of the Galactic plane were used for fitting. 
We adopted Gaussian profiles because they are convenient fitting functions and they were able to reproduce the profiles adequately (see \cite{Reach92, Reach97}).
Examples of five-Gaussian fits are shown in figure \ref{fig:latprofile}.

\begin{figure}
 \begin{center}
  \includegraphics[width=10cm]{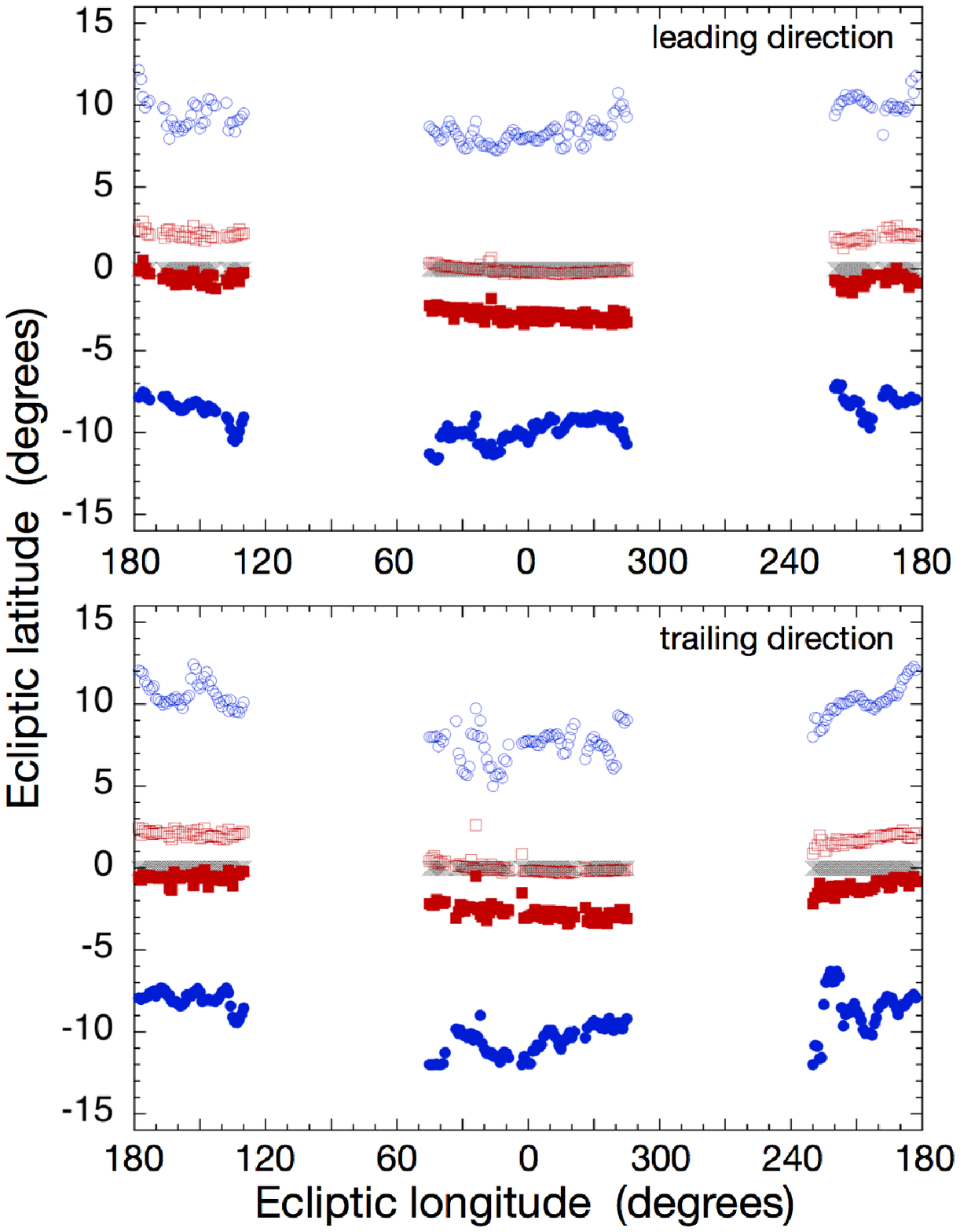} %
 \end{center}
\caption{Ecliptic latitudes of the dust-band peak brightnesses in the Earth-leading (top) and trailing directions (bottom) as a function of geocentric ecliptic longitude along the line of sight. The peak positions of the $\pm\timeform{1.4D}/\timeform{2.1D}$ band are shown in red, the $\pm \timeform{10D}$ band in blue and the circumsolar ring in black. Regions around $45\degree < |\lambda| < 135\degree$ were not used for fitting because of the strong emission near the Galactic plane ($|b| < 15\degree$).}
\label{fig:peak}
\end{figure}

Table \ref{tab:gaussian} summarizes the mean peak surface brightness, mean geocentric ecliptic latitude and band widths obtained from the Gaussian fits. Figure \ref{fig:peak} shows the best-fitting results for the peak positions in the Earth-leading and trailing directions.
There is an apparent annual variation of the band latitudes with geocentric ecliptic longitude.
Sinusoidal variations can be seen in the northern and southern components of each band pair.
The annual variation is caused by the Earth's changing altitude with respect to the midplane of the band pairs.

The sinusoidal variation of the latitude of the dust bands can be expressed using the following parameters: 
(i) the radius of the band pair centred on the Sun (the heliocentric distance), $R$; (ii) the angle of inclination with respect to the ecliptic plane, $i$; (iii) the longitude of the ascending nodes, $\Omega$; (iv) the separation between band pairs, $2z$; and (v) the solar elongation angle of the line of sight, $\epsilon$: see Equations (1) and (2) of \citet{Reach92}.
For AKARI observations at constant solar elongation, $\epsilon \sim 90\degree$, the average latitude of the northern and southern bands (the midplane of the band pairs) is 
\begin{equation}
\tan\beta = -\frac{R}{\sqrt{R^2-1}}~\tan i~\sin\left(\lambda-\Omega\pm\sin^{-1}\frac{1}{R}\right),
\end{equation}
where $\lambda$ is the geocentric ecliptic longitude of the line of sight, and the plus/minus signs in the $\pm$ notation refer to the Earth-leading/trailing direction: see Equation [3] of \citet{Reach92}.
Generally, the heliocentric distance of the dust band ($R$) is determined as a parallactic distance from the band-pair latitudes and the solar elongation angle (\cite{Reach92, Reach97, Spiesman95}). 
Since most AKARI scans were limited to a constant solar elongation, $\epsilon \sim 90\degree$, it is difficult to determine the parallactic distance based on the AKARI data. 
Instead, the phase difference, $\Delta\phi$, between the annual variation of the average latitude of the band pair for scans leading and trailing the Earth can be used to measure the heliocentric distance of the bands, $R$ \citep{Reach91}:
\begin{equation}
\Delta\phi = 2~\tan^{-1}~\sqrt{R^2-1}.
\end{equation}
Once we obtain the heliocentric distance of the band pair ($R$), we can derive the inclination, $i$, and the ascending node, $\Omega$. 
The heliocentric distances of the band pairs, the inclination angles of the dust bands and the locations of the ascending nodes derived from the AKARI maps are listed in Table \ref{tab:geometry}.
%
%
The heliocentric distances of the band pairs are 1.86 and 2.16 a.u., the inclination angles of the dust bands are $\timeform{1.0D}\pm\timeform{0.1D}$ and $\timeform{1.3D}\pm\timeform{0.1D}$, the ascending nodes are located at $112\pm6\degree$ and $95\pm5\degree$ for the $\pm\timeform{1.4D}/\timeform{2.1D}$ and the $\pm\timeform{10D}$ bands, respectively.

\begin{table}
  \caption{Gaussian properties of the dust bands.}\label{tab:gaussian}
  \begin{center}
    \begin{tabular}{llrr}
      \hline
      \hline
      Band  & & Leading & Trailing  \\
      \hline
      \multicolumn{4}{c}{Peak surface brightness (MJy sr$^{-1}$)} \\
      \hline
	  Circumsolar ring & & $2.0\pm0.1$ & $2.2\pm0.1$ \\
      $\pm\timeform{1.4D}/\timeform{2.1D}$ band & North & $0.9\pm0.1$ & $1.0\pm0.1$ \\
                                                 & South & $0.7\pm0.1$ & $0.8\pm0.1$ \\
      $\pm\timeform{10D}$ band & North & $0.6\pm0.1$ & $0.6\pm0.1$ \\
                                                & South & $0.8\pm0.1$ & $0.8\pm0.1$ \\
      \hline
      \multicolumn{4}{c}{Geocentric latitude of the peak (degrees)} \\
      \hline
	  Circumsolar ring & & $0.0\pm0.1$ & $0.0\pm0.1$ \\
      $\pm\timeform{1.4D}/\timeform{2.1D}$ band & North & $1.4\pm0.1$ & $1.4\pm0.1$ \\
                                                 & South & $-1.4\pm0.1$ & $-1.4\pm0.1$ \\
      $\pm\timeform{10D}$ band & North & $9.2\pm0.1$ & $9.3\pm0.1$ \\
                                                & South & $-9.2\pm0.1$ & $-9.3\pm0.1$ \\
      \hline
      \multicolumn{4}{c}{Full width at half maximum (degrees)} \\
      \hline
	  Circumsolar ring & & $32.2\pm0.2$ & $33.0\pm0.2$ \\
      $\pm\timeform{1.4D}/\timeform{2.1D}$ band & North & $4.8\pm0.1$ & $4.4\pm0.1$ \\
                                                 & South & $5.2\pm0.1$ & $4.8\pm0.1$ \\
      $\pm\timeform{10D}$ band & North & $4.8\pm0.1$ & $5.2\pm0.1$ \\
                                                & South & $5.6\pm0.1$ & $5.6\pm0.1$ \\
      \hline
    \end{tabular}
  \end{center}
\end{table}

\begin{table}
  \caption{Geometry of the dust bands in the AKARI maps.}\label{tab:geometry}
  \begin{center}
    \begin{tabular}{lcccl}
      \hline
      Band & $i$ (degrees) & $\Omega$ (degrees) & $R$ (a.u.) & Associated asteroid family \\
     \hline
      $\pm\timeform{1.4D}/\timeform{2.1D}$ band & $1.0\pm0.1$ & $112\pm6$ & 1.86 & Beagle/Karin families \\
      $\pm\timeform{10D}$ band & $1.3\pm0.1$ & $95\pm5$ & 2.16 & Veritas family \\
      \hline
    \end{tabular}
  \end{center}
\end{table}

\begin{figure}
   \begin{center}
    \includegraphics[width=17cm]{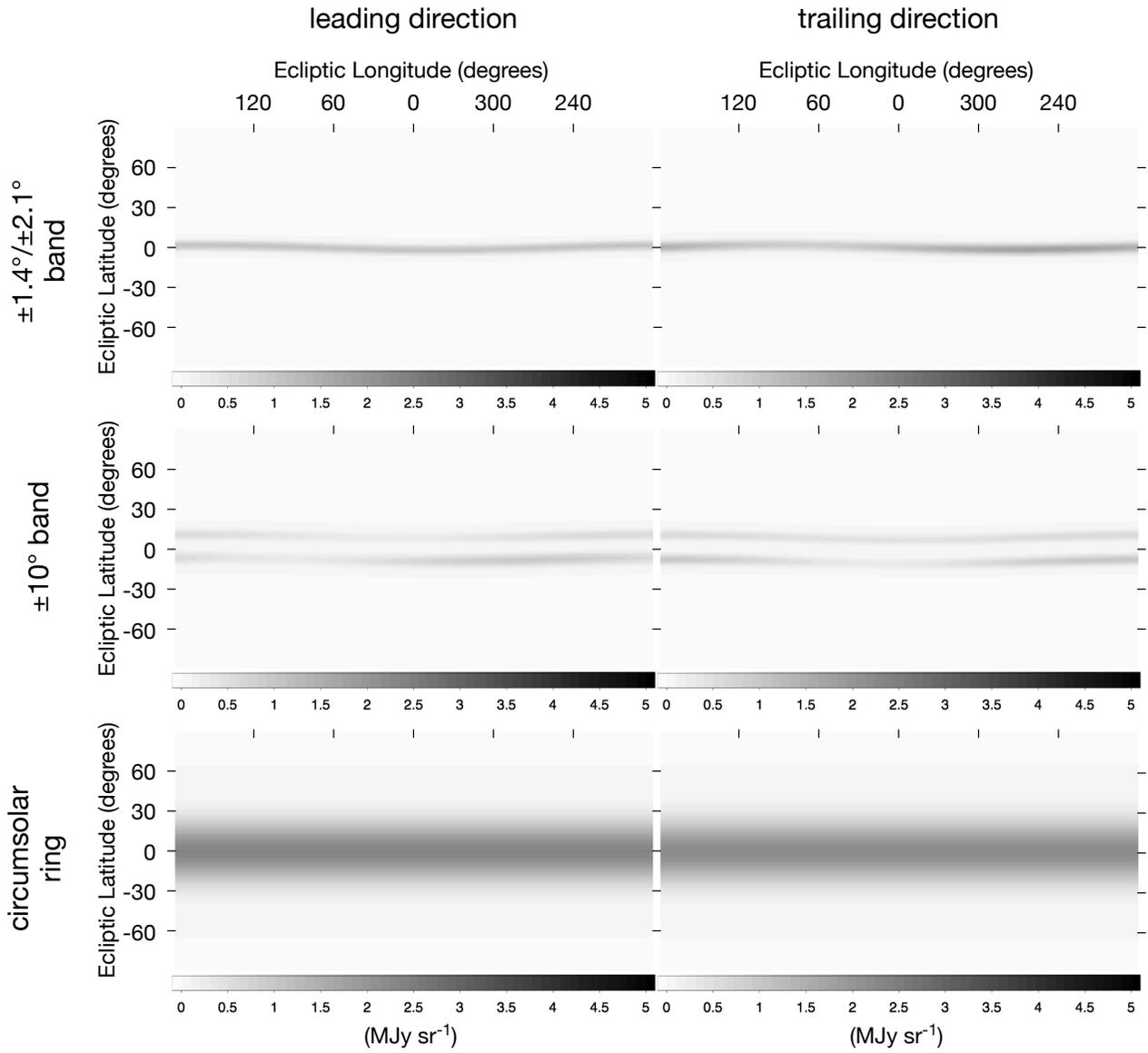} %
   \end{center}
\caption{Model maps of the dust-band components in the AKARI far-IR maps in the Earth-leading (left-hand panels) and trailing (right-hand panels) directions. From top to bottom: $\pm \timeform{1.4D}/\timeform{2.1D}$ band, $\pm \timeform{10D}$ band, circumsolar ring component. Phase difference between northern and southern components can be seen for the $\pm \timeform{10D}$ band. Greyscale as in figure \ref{fig:ltmaps}.}
\label{fig:ZEmodel_lt}
\end{figure}

It is also evident that the surface brightness of the dust bands varies as a function of ecliptic longitude.
\citet{Reach97} refer to the phenomenon that the northern component is bright when the southern component is faint, and {\it vice versa}, as ``glints''.
In the DIRBE maps, the northern and southern glints occur roughly 180$\degree$ apart, while the leading and trailing maps are roughly 180$\degree$ out of phase \citep{Reach97}.
Although the difference between the leading and trailing direction is faint, similar dust-band glints are detected in the AKARI maps, and can be seen in the component maps (figure \ref{fig:ZEmodel_lt}) in particular for the $\pm 10\degree$ band.
We also performed sinusoidal fits to the longitudinal brightness variations of the northern and southern components.

Based on the brightnesses, latitudes and band widths derived from the sinusoidal fitting procedure outlined above (see Tables \ref{tab:gaussian} and \ref{tab:geometry}, figure \ref{fig:peak}), we constructed a model map for the dust-band structure in the far-IR regime. 
Our model maps for each component in the Earth-leading and trailing directions are shown in figure \ref{fig:ZEmodel_lt}.

\subsection{AKARI ZE template maps}

Since these component model maps were derived from the ($I_\mathrm{WIDE-S}-0.30 \times I_\mathrm{WIDE-L}$) difference map,
the surface brightness of the model map is fainter than the real brightness of the dust bands in the WIDE-S map.
To construct dust-band template maps for the four FIS bands, we considered the temperatures of the dust-band components.
As derived in section 3.1, the $\pm \timeform{1.4D}/\timeform{2.1D}$ and $\pm \timeform{10D}$ dust bands have radii of 1.86 a.u. and 2.16 a.u., respectively.
It can thus be expected that the $\pm \timeform{1.4D}/\timeform{2.1D}$ and $\pm \timeform{10D}$ components have typical temperatures of $\sim 200$~K and $\sim 185$~K, respectively.
We assume a temperature for the circumsolar ring component of 278~K, because the circumsolar ring is expected to be composed of dust in the Earth's orbit (1~a.u.).
If we assume that the dust temperatures of the dust bands are $T_{\rm d} \sim 185$~K, 200~K, and 278~K, their brightnesses relative to those of the WIDE-S band data are 1.376, 1.390 and 1.439 for the N60 band, 0.212, 0.208 and 0.193 for the WIDE-L band, and 0.133, 0.130 and 0.119 for the N160 band, respectively. 
To construct the corresponding dust-band maps in the four FIS bands for the circumsolar-ring component ($T=278$~K), we multiplied the brightness of the component model by the following factors:
\begin{itemize}
\item N60: $1.439/(1.0-0.30 \times 0.193) = 1.53$;
\item WIDE-S: $1.0/(1.0-0.30 \times 0.193) = 1.06$;
\item WIDE-L: $0.193/(1.0-0.30 \times 0.193) = 0.20$;
\item N160: $0.119/(1.0-0.30 \times 0.193) = 0.13$.
\end{itemize}
The $\pm \timeform{1.4D}/\timeform{2.1D}$ and $\pm \timeform{10D}$ band-pair maps were derived similarly, and we constructed the AKARI dust-band templates using these component maps as follows.

\begin{figure}
 \begin{center}
  \includegraphics[width=17cm]{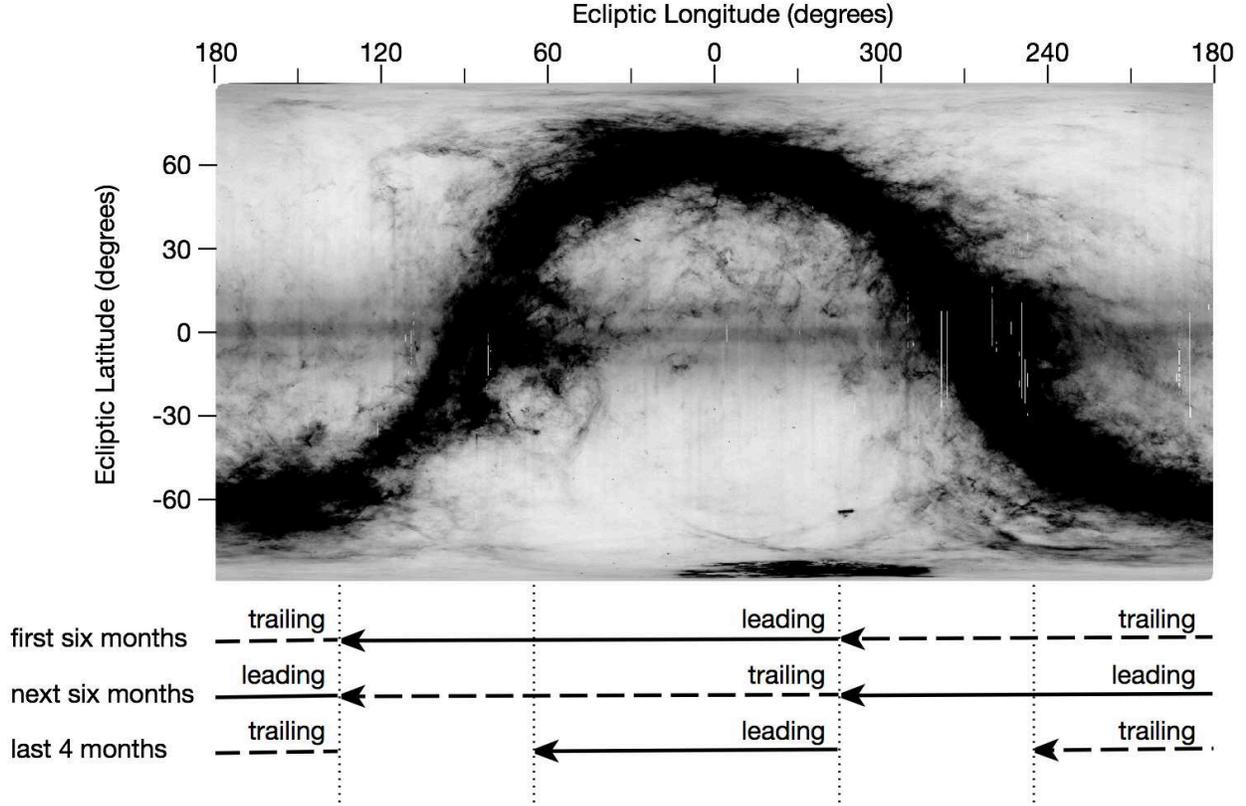} %
 \end{center}
\caption{All-sky map in the WIDE-S band, in ecliptic coordinates, showing which of the leading and trailing scans cover the relevant sky regions. See text for more details.
}
\label{fig:season}
\end{figure}

The publicly released AKARI maps have been processed with data in both the leading and trailing directions obtained during different semi-annual seasons.
To construct the ZE template of the dust bands for the AKARI maps, we need to take into account the numbers of observations of the leading and the trailing scans for each pixel.
Figure \ref{fig:season} shows which of the leading and trailing scans cover the relevant sky regions.
During the first 6 months, the leading scans observed the $315\degree < \lambda < 360\degree$ and $0\degree \leqq \lambda < 135\degree$ regions, while the trailing scans covered $135\degree < \lambda < 315\degree$.
Over the next 6 months, $135\degree < \lambda < 315\degree$ was covered by the leading scans, and $315\degree < \lambda < 360\degree$ and $0\degree \leqq \lambda < 135\degree$ by the trailing scans.
During the last 4 months, the leading scans observed the $315\degree < \lambda < 360\degree$ and $0\degree \leqq \lambda < 65\degree$ regions, while $135\degree < \lambda < 245\degree$ was covered by the trailing scans (see figure \ref{fig:season}).

\begin{figure}
 \begin{center}
   \includegraphics[width=10cm]{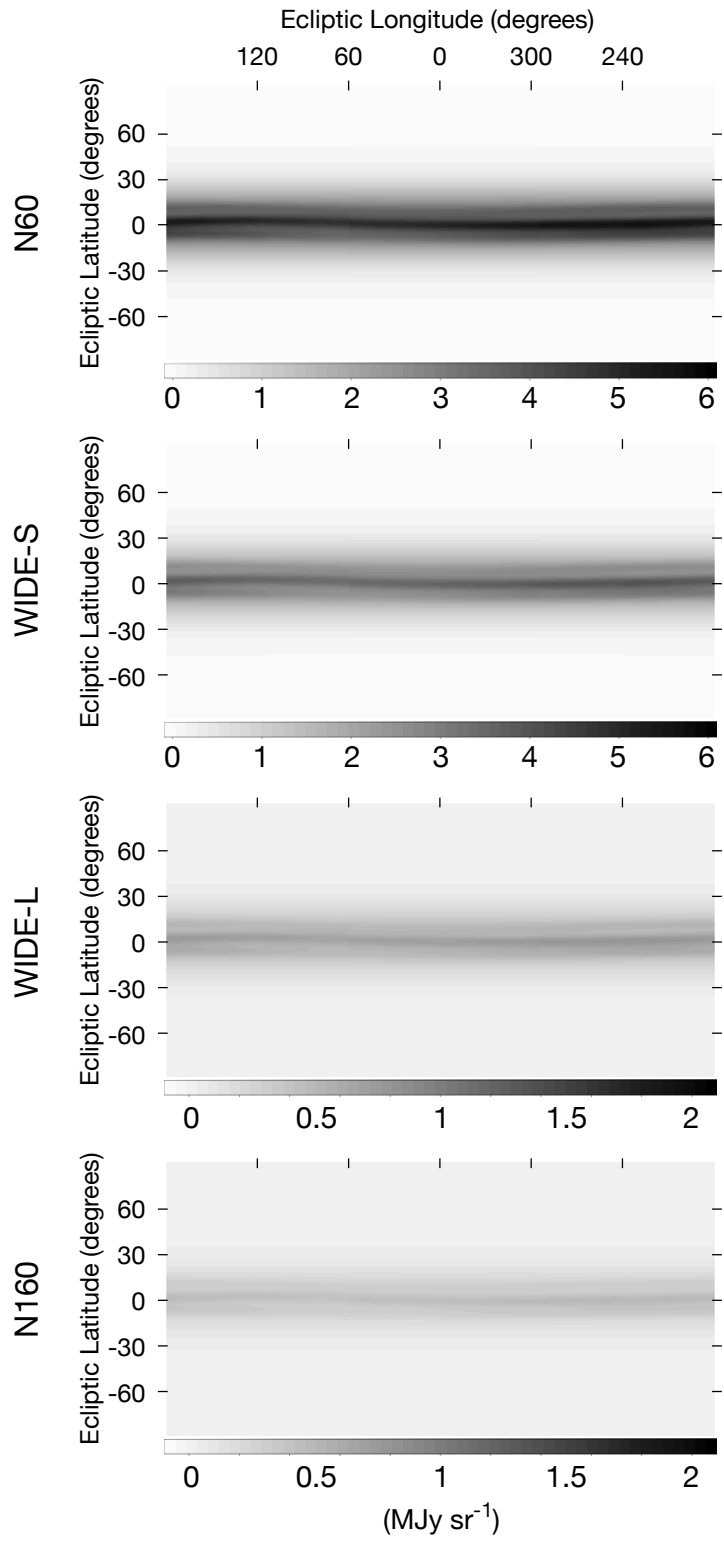} %
 \end{center}
\caption{Dust-band template maps based on the AKARI far-IR maps. From top to bottom: N60, WIDE-S, WIDE-L and N160. Greyscale as in figure \ref{fig:ltmaps}.}
\label{fig:ZEmodel_allsky}
\end{figure}

Based on the conditions discussed above, we combined leading (L) and trailing (T) dust-band maps to construct the AKARI ZE dust-band template, as follows:
\begin{itemize}
\item $0\degree \leqq \lambda < 65\degree$: (2L+T)/3;
\item $65\degree \leqq \lambda < 135\degree$: (L+T)/2;
\item $135\degree \leqq \lambda < 245\degree$: (L+2T)/3;
\item $245\degree \leqq \lambda < 295\degree$: (L+T)/2;
\item $295\degree \leqq \lambda < 360\degree$: (2L+T)/3.
\end{itemize}
The resulting dust-band templates for the AKARI maps in the four FIS bands are shown in figure \ref{fig:ZEmodel_allsky}.

\begin{figure}
  \begin{center}
    \includegraphics[width=17cm]{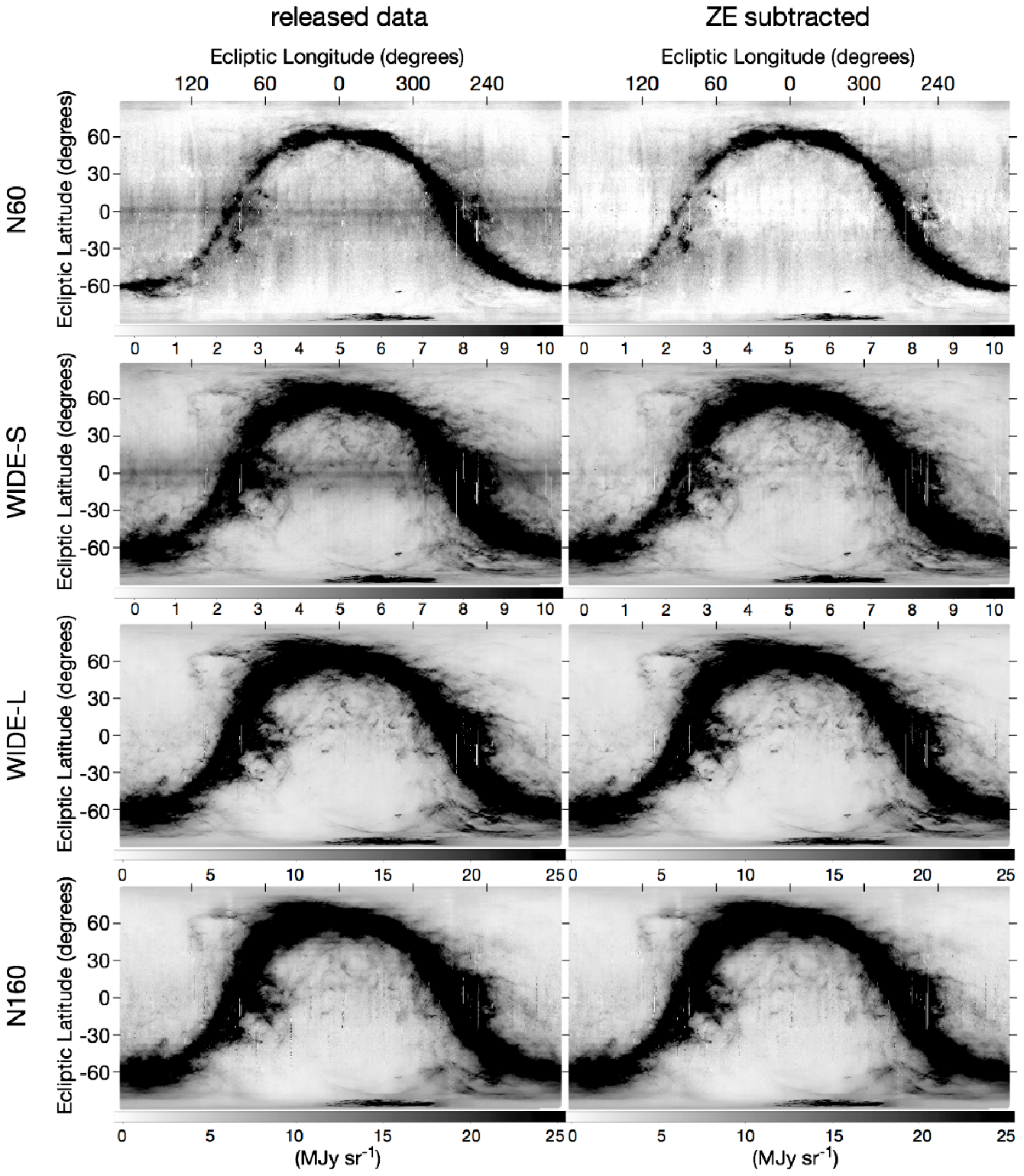} %
  \end{center}
  \caption{AKARI far-IR all-sky surface brightness maps constructed based on the publicly released data (left-hand panels) and ZE-subtracted maps (right-hand panels). From top to bottom: N60, WIDE-S, WIDE-L and N160. Data in both the leading and the trailing directions were used to construct both maps. Greyscale as in figure \ref{fig:ltmaps}.}
\label{fig:zodisub}
\end{figure}

%
%
The AKARI dust-band template thus obtained was subtracted from the all-sky map, for each band (see figure \ref{fig:zodisub}).
We can reduce the residual contribution of the ZE to $\lesssim 0.5$~MJy~sr$^{-1}$ near the ecliptic plane by
subtracting these dust-band maps from the AKARI far-IR all-sky maps (see figure \ref{fig:zodisub} and figure \ref{fig:latprofile}).

\section{Discussion}

\subsection{Dust-band structures}

In previous studies, structure in the dust bands was mainly discussed based on IRAS and DIRBE data at 25~$\micron$,
because the band structure, in particular of the $\pm \timeform{10D}$ dust band, is faint at wavelengths longer than 60~$\micron$ \citep{Sykes88, Spiesman95, Reach97}.
In the AKARI maps, we can detect the $\pm \timeform{10D}$ band structure much more clearly in the 65~$\micron$ and 90~$\micron$ bands, compared with the DIRBE maps.
We derived the heliocentric distances of dust bands from the phase difference based on the AKARI 90~$\micron$ map: 1.86 a.u. and 2.16 a.u. for the $\pm\timeform{1.4D}/\timeform{2.1D}$ and the $\pm\timeform{10D}$ bands, respectively.
\citet{Reach92} found the heliocentric distances of $1.32 \pm 0.05$ a.u. and $2.04 \pm 0.04$ a.u. using IRAS data. 
Based on the DIRBE data, \citet{Spiesman95} derived the parallactic distances of $1.37 \pm 0.17$ a.u. and $2.05 \pm 0.13$ a.u., while \citet{Reach97} found $1.7 \pm 0.2$ a.u. and $2.4 \pm 0.3$ a.u., respectively.
Although the heliocentric distances estimated with the AKARI far-IR data are consistent with that of previous studies, AKARI results of the $\pm\timeform{1.4D}/\timeform{2.1D}$ band is slightly larger than those derived in previous studies.
\citet{Nesvorny06a} made the simulations of orbital evolution for the dust particles released from the Karin and Veritas family. 
They estimated the relative contribution to the total IR surface brightness from particles at various distances, and 
suggest that the contribution of particles at further than 2 a.u. is stronger in the IRAS 60~$\micron$ than that at shorter wavelengths. 
They also found that large particles ($\sim 500~\micron$) interact with strong secular resonances at the heliocentric distance $\approx 2$~a.u., while rapidly drifting small particles only experience modest perturbations.
This suggests that the larger grains at $> 2$~a.u. are dominant for the $\pm\timeform{10D}$ band, while smaller dust grains at $< 2$~a.u contribute to the $\pm \timeform{1.4D}/\timeform{2.1D}$ dust band.
Since AKARI results are mainly based on the 90~$\micron$ and 140~$\micron$ band data, large dust particles at further heliocentric distances ($> 2$~a.u.) contribute to the AKARI results more significantly than that 
in the IRAS results.

We fitted the $\pm \timeform{1.4D}/\timeform{2.1D}$ band with a single Gaussian function. Weak but sharp structures ($< 0.5$~MJy~sr$^{-1}$ on scales of a few degrees) remaining near the ecliptic plane in the ZE-subtracted N60 and WIDE-S maps (see figure \ref{fig:latprofile} and figure \ref{fig:zodisub}) can probably be attributed to the $\pm\timeform{2.1D}$ band, because the separation of this band pair is $\sim 4\degree$. 
It has been suggested that the $\pm \timeform{1.4D}$ band originates from the breakup of the Beagle asteroid family $< 10$ Myr ago, while the $\pm \timeform{2.1D}$ band may have come from the Karin asteroid family, which formed around 5.8 Myr ago \citep{Nesvorny03, Nesvorny08}.
It is likely that a broad $\pm \timeform{1.4D}$ band can be mainly attributed to the older Beagle family and that the $\pm\timeform{2.1D}$ dust band originating from the recent breakup of the Karin family still exhibits sharp, well-defined structures.

\begin{figure}
 \begin{center}
   \includegraphics[width=17cm]{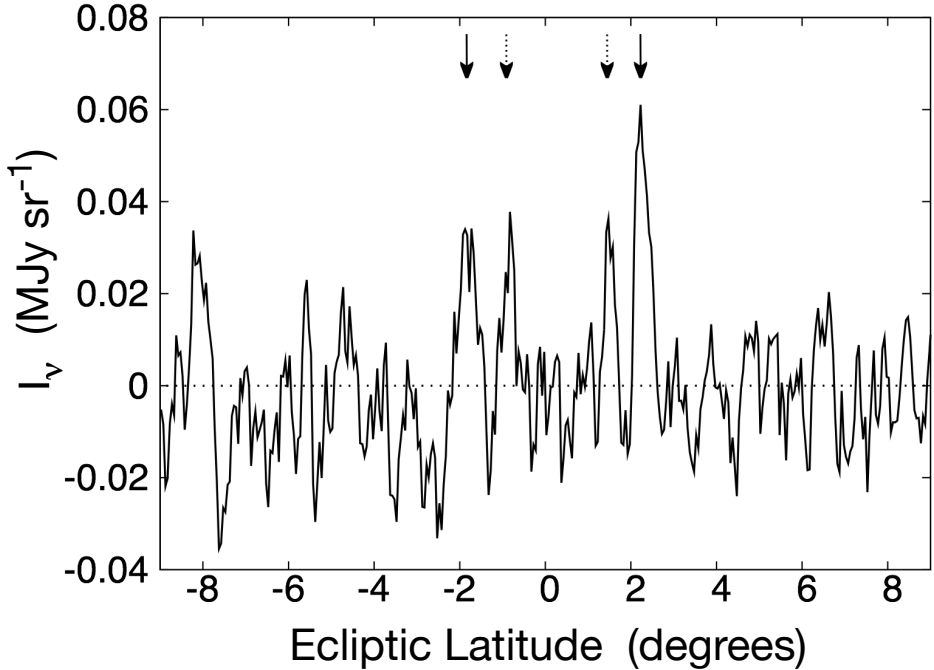} %
 \end{center}
\caption{Spatially filtered latitudinal profile of the region around $\lambda = 125\degree$ in the ``difference map'' for the leading direction. Solid and dotted arrows indicate the $\pm \timeform{2.1D}$ band and the $\pm \timeform{1.4D}$ band, respectively. High-resolution AKARI data has a capability to detect these fine-scale structures.}
\label{fig:hi-res_profile.eps}
\end{figure}

To see the fine structure of these two $\pm \timeform{1.4D}$ and $\pm \timeform{2.1D}$ bands,
we created the ``difference map'' image at near the ecliptic plane with the $\timeform{0.05D}$ resolution. 
Similar to figure \ref{fig:latprofile}, the image is centered at ecliptic longitude $\lambda = 125\degree$ and latitude $\beta = 0\degree$, covering a $10\degree \times 20\degree$ area in the leading direction, because this region is free of significant galactic dust emission.
The image is spatially filtered by removing a boxcar average of 1$\degree$ width in the similar way to \citet{Reach97} and \citet{Spiesman95}, and we constructed a profile of the surface brightness as a function of ecliptic latitude, coadding in 10$\degree$ longitude ranges.
The latitudinal profile is shown in figure \ref{fig:hi-res_profile.eps}. 
The profile shows four individual peaks that are also seen in the figure 9 based on the IRAS 25~$\micron$ data in \citet{Reach97}.
This result demonstrates the advantage of the AKARI data for the investigation of the finer-scale dust band structures in the far-IR regime.

In addition to the major bands at $\pm \timeform{1.4D}/\timeform{2.1D}$ and $\pm 10\degree$, a weak $\pm 17\degree$ band pair, and potential $\pm 6\degree$ and $\pm 13\degree$ band pairs have been suggested in previous studies \citep{Sykes88, Reach97, Ishiguro99a}. 
These fainter band pairs are not apparent in the AKARI maps, although the $\pm 17\degree$ band pair may have been detected (see also figure \ref{fig:latprofile}).
These bands will be discussed based on higher-resolution maps and Fourier-filtering techniques in a forthcoming paper. 
In the present analysis we simply adopted Gaussian profiles for the fitting, although a Gaussian shape is not theoretically motivated.
To improve the dust-band template maps, we will consider a more realistic model for the dust bands that has been proposed on physical grounds (e.g., \cite{Sykes86, Reach97}) in future analysis.

The dust properties of the asteroidal dust bands at wavelengths longer than 100~$\micron$ remain poorly known
because the ZE brightness rapidly becomes faint at longer far-IR wavelengths. 
IRAS was not equipped with bands longer than 100~$\micron$. 
Although DIRBE carried 140~$\micron$ and 240~$\micron$ bands, the spatial resolution and the data quality of these bands were 
not good 
and no dust bands were identified at the DIRBE 140~$\micron$ or 240~$\micron$ bands \citep{Spiesman95}.
Although AKARI is equipped with WIDE-L (140~$\micron$) and N160 (160~$\micron$) bands with a higher spatial resolution than DIRBE, the asteroidal dust bands apparently cannot be identified in either the 140~$\micron$ or the 160~$\micron$ images.
Based on data from DIRBE and the Far Infrared Absolute Spectrophotometer (FIRAS) instrument onboard COBE, it has been suggested that the ZE spectrum exhibits a break at $\lambda \sim 150~\micron$ and that the emissivity becomes wavelength-dependent, falling off as $\lambda^{-2}$, at wavelengths longer than 150~$\micron$ \citep{Fixen02}.
In figure 2 of \citet{Fixen02}, the emissivity of the ZE spectrum actually decreases from $\sim 100~\micron$ onwards.
Non-detection of asteroidal dust bands in the AKARI WIDE-L and N160 bands may support the emissivity decrease in the longer-wavelength regime.

The peak positions of the circumsolar-ring component are naturally almost all located near the ecliptic plane. They do not show any clear sinusoidal change, because AKARI observed the dust in the circumsolar ring near the Earth's orbit (see figure \ref{fig:peak}).
IRAS and COBE/DIRBE confirmed that the circumsolar-ring component is brighter in the direction trailing the Earth than in the leading direction \citep{Dermott94, Reach95}.
\citet{Dermott94} suggested that the ``total brightness'' trailing the Earth was greater than that leading the Earth by $\sim 3$--4\% in the 12--60~$\micron$ bands.
In the AKARI difference map based on the 90~$\micron$ and 140~$\micron$ bands, we can see the trailing--leading asymmetry, but the asymmetry in the ``total ZE brightness'' is less than 3\%. (Keep in mind that the smooth cloud component of the ZE has already been subtracted from the AKARI maps.) It is likely that the IRAS 12--60~$\micron$ bands observed a region closer to the Earth and of higher density than that probed by the AKARI 90~$\micron$ band.

\subsection{Data quality of the AKARI ZE-subtracted maps}

\begin{figure}
 \begin{center}
  \includegraphics[width=17cm]{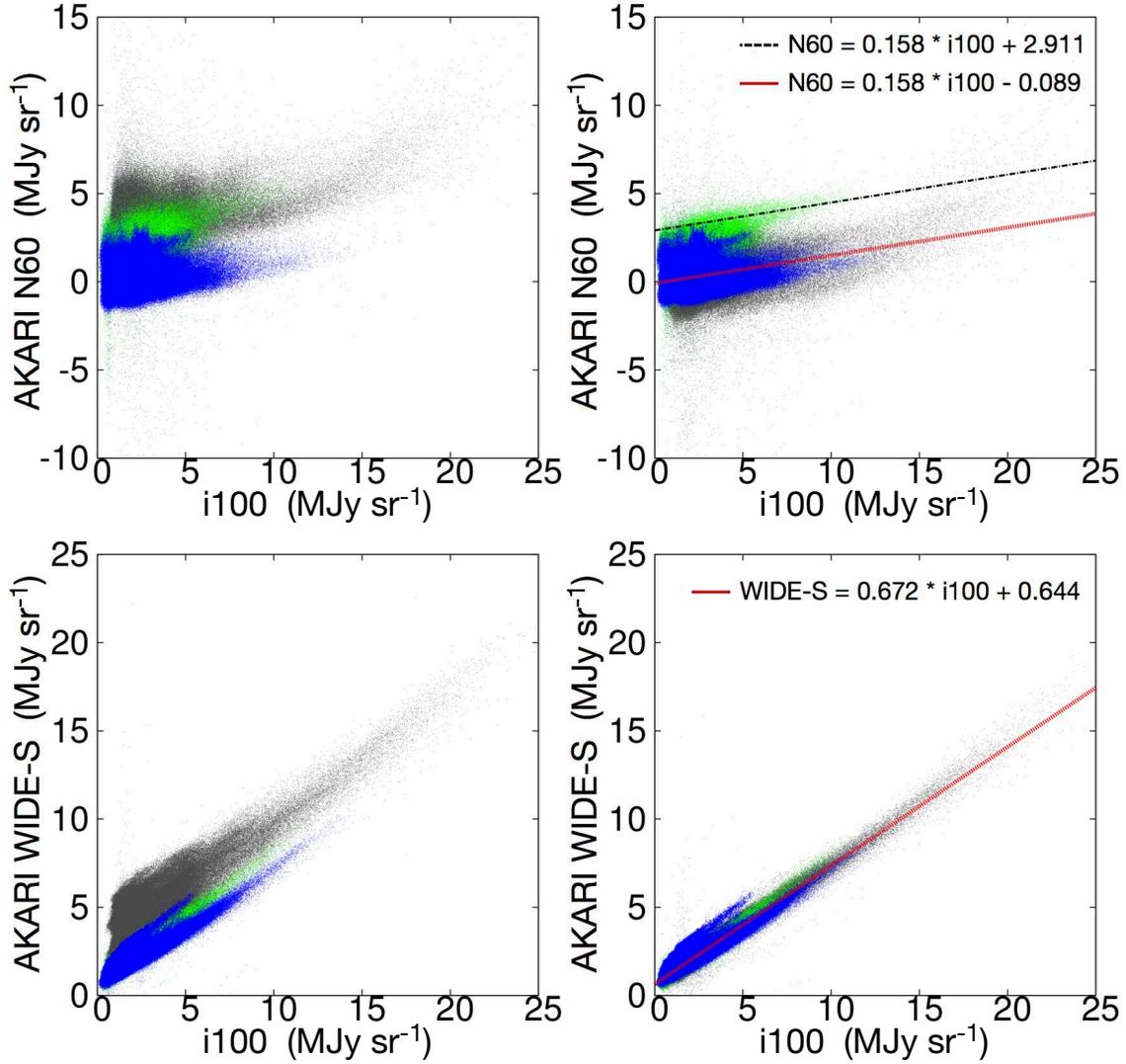} %
 \end{center}
\caption{Scatter plots of the AKARI N60 (top) and WIDE-S (bottom) data vs. Galactic 100~$\micron$ emission \citep{SFD98} in three ecliptic latitude regions at $0\degree < |\beta| < 30\degree$ (gray), $30\degree < |\beta| < 50\degree$ (green) and $50\degree < |\beta| < 90\degree$ (blue).
Left: publicly released AKARI all-sky survey data before ZE subtraction. 
Right: ZE-subtracted AKARI data.
Red lines show linear fits to the ZE-subtracted AKARI intensity as a function of i100. The mid-latitude data have been excluded from the fits to the N60-band observations. For the WIDE-S band, all data have been used for the fitting. The black line in the N60-band panel represents the intensity obtained by adding 3~MJy~sr$^{-1}$ to the best-fitting results for the low- and high-latitude regions.}
\label{fig:AKARI-SFD}
\end{figure}

Contributions from small-scale ZE structures remain present in the publicly released AKARI images. The estimated intensities of the residual emission in the AKARI FIR images reach a maximum near the ecliptic plane of $< 5$~MJy~sr$^{-1}$ and $< 4$~MJy~sr$^{-1}$
for the N60 and WIDE-S
bands, respectively \citep{Doi15, Takita15}.
Using the AKARI dust-band template maps derived in this paper, we can reduce the residual ZE contribution
in the AKARI far-IR all-sky survey maps to $\lesssim 0.5$~MJy~sr$^{-1}$ near the ecliptic plane.
Since interstellar dust emission mainly contributes to the ZE-subtracted far-IR maps, we compared the ZE-subtracted AKARI data in the N60 and WIDE-S bands with the Galactic dust emission data.
We do not consider the WIDE-L and N160 bands here, because the contribution of the ZE is negligible for these bands. 
We used the Galactic 100~$\micron$ emission map of \citet{SFD98}, which we refer to as i100, because it was constructed using far-IR data from IRAS and DIRBE, and it is currently the most widely used dust emission for a variety of astronomical studies.
Figure \ref{fig:AKARI-SFD} shows a pixel-to-pixel ($\timeform{0.2D} \times \timeform{0.2D}$) comparison
of the AKARI intensity and the i100 map at high Galactic latitudes ($|b| > 25\degree$). 
We use the i100 map 
without any modification considering the difference in wavelengths. 
For comparison, we divide the sky into three ecliptic latitude regions: $0\degree < |\beta| < 30\degree$ (low latitude), $30\degree < |\beta| < 50\degree$ (mid latitude) and $50\degree < |\beta| < 90\degree$ (high latitude). The asteroidal dust bands and the circumsolar-ring components mainly contribute to the low-latitude region, while the high-latitude region ($|\beta| > 50\degree$) is negligibly affected by the dust bands. 

Before subtracting the dust-band components, an excess can be seen owing to dust-band emission in the low-latitude region in both the N60 and WIDE-S bands (see the left-hand panels of figure \ref{fig:AKARI-SFD}).
The mid-latitude region in the N60 band also shows some excess compared with the high-latitude region.
After subtraction of the dust-band templates, all regions exhibit similar trends and show a good correlation with the i100 surface brightness,
except for the mid-latitude region in the N60 band, which still exhibits some excess (see the right-hand panels of figure \ref{fig:AKARI-SFD}).
For the intensity uncertainties we use the standard deviations with respect to the i100 intensities based on the derived correlations for various intensity ranges in the similar way to \citet{Takita15}, $0.158 \times I_\mathrm{i100} - 0.089$ for N60 and $0.672 \times I_\mathrm{i100} + 0.644$ for WIDE-S.
The standard deviations are 23.0\% at 1--2.5~MJy~sr$^{-1}$, 20.4\% at 2.5--4.0~MJy~sr$^{-1}$ and 20.2\% at 4.0--6.3~MJy~sr$^{-1}$ for N60 at low and high latitudes, while they are 15.1\% at 1--2.5~MJy~sr$^{-1}$, 13.6\% at 2.5--4.0~MJy~sr$^{-1}$ and 9.7\% at 4.0--6.3~MJy~sr$^{-1}$ for the WIDE-S band for the all-sky data.
We note that these estimates pertain to a spatial resolution of $\timeform{0.2D}$, although \citet{Takita15} derived uncertainties based on the DIRBE intensity with resolution of $\timeform{0.7D}$, excluding the region $|\beta| < 40\degree$. 
We can achieve an equivalent level of flux calibration for the region at $< 10$~MJy~sr$^{-1}$ to \citet{Takita15} even for $|\beta| < 40\degree$ using the ZE-subtraction method developed in this paper.


The residual intensity in the mid-latitude region remaining in the N60 band can also be seen in the all-sky map (figure \ref{fig:zodisub}).
The residual intensity at mid latitudes in the N60 band is $\sim 3$~MJy~sr$^{-1}$ (see figure \ref{fig:AKARI-SFD}).
This may be attributed to the residuals after subtraction of the smooth cloud component using the DIRBE ZE model.
The DIRBE ZE model may be incomplete for use with the AKARI data, even for the smooth cloud component. 

\section{Summary}

We have investigated the geometry of the zodiacal dust bands in the AKARI far-IR maps and constructed template maps of the asteroidal dust bands and the circumsolar-ring components based on the AKARI far-IR all-sky survey maps. 
In the AKARI maps, we can clearly detect the $\pm \timeform{10D}$ band structure as well as that of the $\pm\timeform{1.4D}/\timeform{2.1D}$ band in the 65~$\micron$ and 90~$\micron$ bands.
In addition to the major bands at $\pm \timeform{1.4D}/\timeform{2.1D}$ and $\pm 10\degree$, a weak $\pm 17\degree$ band pair, and potential $\pm 6\degree$ and $\pm 13\degree$ band pairs have been suggested in previous studies.
These fainter band pairs are not apparent in the AKARI maps, although a possible $\pm 17\degree$ band pair may have been detected.
No evident dust-band structures are identified in the 140~$\mu$m and 160~$\mu$m bands.
Applying five-Gaussian fits to the difference map composed of the 90~$\micron$ and 140~$\micron$ bands, we constructed dust-band template maps for the four AKARI far-IR bands.   
By subtracting the dust-band templates derived here, we can obtain ZE-subtracted AKARI far-IR all-sky maps.
With the ZE-subtracted maps, we can achieve an equivalent level of flux calibration to that of \citet{Takita15}, even for the regions at $|\beta| < 40\degree$.

\bigskip

\begin{ack}

This research is based on observations with AKARI, a JAXA project with the participation of ESA.
This work has been supported by JSPS KAKENHI Grant Number 19204020, 21111005, 25247016, 25400220,
and 15J10278. 
We thank M. Ishiguro for the careful reading and inspiring comments on this work.
We also thank T. Kondo and D. Ishihara for the useful discussions and suggestions.

\end{ack}


\end{document}